\documentclass[11pt,dutch,titlepage]{article}%
\usepackage{amssymb}
\usepackage{float}
\usepackage{makeidx}
\usepackage{graphicx}
\usepackage{amsmath}
\usepackage{amsfonts}
 \usepackage{tikz}
\usepackage[a4paper]{geometry}

\newcommand{\bpm}{\begin{pmatrix}}
\newcommand{\epm}{\end{pmatrix}}

\begin{document}

\baselineskip=16.pt
\newcounter{ctr}

\begin{center}
{\Large {\bf Tangent Developable Orbit Space of an Octupole}}\\
\vspace{1cm}
Jan-Peter B{\" o}rnsen, Anton E. M. van de Ven\\
{\it University College Utrecht}\\
(Dated: \today)\\
\vspace{0.5cm}
jan-peter@boernsen.net \quad A.E.M.vandeVen@uu.nl
\end{center}
\vspace{1cm}
{\bf Abstract}: The orbit space of an octupole, a traceless symmetric third-rank tensor of $SO(3)$, is shown to be a three-dimensional body with three cusps and two cuspidal edges. It is demonstrated that for a unique choice of orbit space coordinates its boundary turns into a tangent developable surface of degree six which we identify with one in the classification by Chasles and Cayley. The close relation of the octupole's orbit space to the moduli space of a binary sextic form is described and relevant work by Clebsch and Bolza from the nineteenth century is recalled. Upon complexification, the octupole's orbit space yields the geometry of the moduli space of  hyperelliptic curves of genus two. Its boundary is found to be a non-orientable tangent developable surface. As preamble, the orbit space of a set of three vectors is shown to be bounded by Cayley's nodal cubic surface and the boundary of the orbit space for a vector plus a quadrupole is found to correspond to  Cremona's first class in his classification of quartic ruled surfaces.
\vspace{2cm}

\section{Introduction}
The geometry of orbit spaces is important in physics, among others in studying spontaneous symmetry breaking in gauge theories \cite{K82}, \cite{DIK}, and in the phenomenological description of phase transitions \cite{TT}. In gauge theories, renormalizability restricts the Higgs potential to terms at most quartic in the fields while in condensed or soft matter physics no such restriction holds for the Ginzburg-Landau potential. Other areas of physics in which the geometry of orbit spaces is of interest include quantum information theory \cite{BZ}, bifurcation theory \cite{CL}, and screw theory \cite{DG}.\\

Systematic, explicit, results for orbit spaces of small, coregular\footnote{A representation is  coregular if there are no polynomial relations among its invariants.}, representations of classical and exceptional Lie groups were obtained in \cite{K} with earlier work in \cite{MR} and \cite{Li}. We also mention the work of J{\'a}ric \cite{J} who found 2D-projections of orbit spaces for the $\ell = 2$, 4 and 6 irreducible representations (irreps) of $SO(3)$, keeping only cubic and quartic invariants. Order parameters with odd values of $\ell$ were generally ignored because they are odd under inversion \cite{SNR} but with the advent of, among others, bent-core liquid crystals \cite{LuR}, spin-3 Bose-Einstein condensates \cite{KU}, and the tetrahedral group as explanation for the neutrino mass matrix \cite{BG}, this has changed. Recently, there is also renewed interest in higher-rank order parameters to describe systems with exotic orientational states \cite{NZ}.\\

With this in mind, the work of \cite{K} and \cite{J} will be extended here to some non-coregular representations of $SO(3)$. In particular, the geometry of the orbit space for the $\ell = 3$ irreps of $SO(3)$, i.e. a traceless symmetric third rank tensor or octupole, will be described in detail. It will be demonstrated that this orbit space is a three-dimensional body whose boundary surface has three cusps, each cusp corresponding to a particular maximal isotropy (or stability) subgroup for this representation. Furthermore, by exploiting the freedom to make homogeneous polynomial redefinitions of the invariants which constitute the Hilbert basis, we shall show that the boundary surface can be turned into the tangent developable of one of the curves which connect the cusps.\\

Due to the well-known relation between $SO(3)$ and $SU(2)$, the invariants of a traceless symmetric rank-$\ell$ tensor of $SO(3)$ are the same as those of an $SU(2)$ binary form of degree $2\ell$. Since such binary forms play a central r\^ole in the description of the moduli space of (hyper)elliptic curves of genus $g = \ell - 1$ \cite{LeRi}, our orbit space for an octupole also yields a geometrical picture of the moduli space of genus two hyperelliptic curves.\\  
% [25-07-2017] 

We will begin by studying the orbit spaces for a set of three vectors and for a vector plus quadrupole tensor. In both cases there exists namely a single polynomial relation which expresses the non-negativity of the square of the unique pseudo-scalar\footnote{In \cite{Po} all representations of $SL(2,C)$ with a single polynomial relation among their invariants were given. We find here the orbit spaces for all such cases which also can be viewed as tensors of $SO(3)$, except for the case of two quartic binary forms, i.e. two $\ell = 2$ irreps of $SO(3)$.}. We will see that this also holds for an $\ell = 3$ tensor so these simpler cases provide a useful introduction to the complicated $\ell = 3$ case. Furthermore, the geometries of these simpler orbit spaces turn out to be interesting in their own right. For a general introduction to orbit spaces we refer to \cite{AS}\footnote{Unlike \cite{K}, whom we follow here, the authors of \cite{AS} do not define dimensionless orbit space coordinates, i.e. they do not normalize on the basic quadratic invariant. Hence, their orbit spaces are non-compact and have one dimension more than ours, making it harder to visualize such spaces.} and \cite{MZ}. A systematic study of orbit spaces appears in \cite{SaVa}, \cite{SaVa2}. A readable text on classical invariant theory is provided by \cite{Ol}. The graphical representations of the orbit spaces in this paper were produced by means of {\it Mathematica}. This program was also indispensable in performing the extensive algebra involved in this investigation.\\ 

As this work neared completion a rather different approach to octupolar order appeared in \cite{CQV}, not based on classical invariant theory but on a generalization of the eigenvalue method to higher rank tensors. No mention is made in \cite{CQV} of the orbit space of an octupole and, at present, the precise relation to our work is not clear.

\section{Simple Examples}

In order to arrive at a description of the orbit space for the $\ell = 3$ irreps we first describe simpler representations whose orbit spaces are also determined by a single polynomial relation among the invariants, a syzygy, namely the non-negativity of the square of a unique pseudo-scalar. 

\subsection{Quadrupole Tensor}
The orbit space of an $\ell = 2$ representation, i.e. a quadrupole tensor (a $3\times 3$ traceless symmetric tensor), is well-known, see e.g. \cite{J}, and plays a prominent role in the description of liquid crystals \cite{GLJ}. It is included here for completeness and to set our notation.\\

Any invariant of a quadrupole tensor $A$ takes the form ${\rm tr} A^n$ with $n>1$. The Cayley-Hamilton theorem tells us that the traceless part of $A^3$ is proportional to $A$
\begin{equation}\label{HC-id}
A^3 -  \frac13 I\ {\rm tr} A^3 = \frac12 A\ {\rm tr} A^2  
\end{equation}
and this implies that a minimal integrity basis, or Hilbert basis, consists of
\begin{equation}
{\rm tr} A^2 \ ,\quad  {\rm tr} A^3 
\end{equation}
Note that ${\rm tr} A^3 = 3 \det A$ and we have the inequality (proven below)
\begin{equation}\label{ineq1}
({\rm tr} A^3)^2 \leq \frac16 ({\rm tr} A^2)^3 
\end{equation}
We assume $A\neq 0$, so ${\rm tr} A^2 > 0$, and define the normalized orbit space coordinate
\begin{equation}\label{z-coordinate}
z = \frac{\sqrt{6}\,{\rm tr} A^3}{\sqrt{{\rm tr} A^2}}\quad , \quad z\in [-1,1]
\end{equation}
Equivalently, we restrict to unit quadrupoles, i.e. we set ${\rm tr} A^2 = 1$ and $z = \sqrt{6}\,{\rm tr} A^3$. 
A convenient parametrization of the diagonal form of such a quadrupole is given by\footnote{This parametrization seems to be not well-known but is commonly used e.g. in nuclear physics.} 
\begin{equation}\label{Adiag}
A = \sqrt\frac23\ {\rm diag}[\cos\alpha_+, \cos\alpha_-, \cos\alpha] 
\quad ,\quad \alpha_\pm \equiv \alpha \pm \frac{2\pi}3
\end{equation}
Note that $\cos\alpha_+\leq\cos\alpha_-\leq\cos\alpha$ for $\alpha\in [0,\frac{\pi}3]$ so the eigenvalues are ordered, and we have the trigonometric properties
\begin{eqnarray}
\cos\alpha_+ \, +\, \cos\alpha_-\, +\, \cos\alpha\ &=& 0\\ 
\cos^2\alpha_+ + \cos^2\alpha_- + \cos^2\alpha &=& \frac32\\
\cos^3\alpha_+ + \cos^3\alpha_- + \cos^3\alpha &=& \frac34 \cos 3\alpha 
\end{eqnarray}
The first two identities imply ${\rm tr} A = 0$ and ${\rm tr} A^2 = 1$ and the last identity informs us that the $\ell =2$ orbit space possesses the simple trigonometric parametrization
\begin{equation}
z = \cos 3\alpha \quad ,\quad \alpha \in [0,\frac{\pi}3]
\end{equation}
This also proves inequality (\ref{ineq1}) which reduces to $\cos^2 3\alpha\leq 1$. Note the special cases
\begin{eqnarray}
\alpha = 0\  &:&  A = \frac1{\sqrt{6}}\,{\rm diag}[-1, -1,  2]\ ,\ \ z = 1\\
\alpha =\frac{\pi}6  &:&  A = \frac1{\sqrt{2}}\,{\rm diag}[-1,\,0,\,1]\ ,\ \ z = 0\\ 
\alpha =\frac{\pi}3 &:& A = \frac1{\sqrt{6}}\,{\rm diag}[-2,\,1,\,1]\ ,\ \ z = -1
\end{eqnarray}
Hence, the $\ell =2$ orbit space $\Omega_2$ is a straight line segment (see Figure 1).
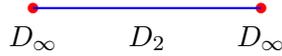
\begin{figure}  
\begin{center} 
\begin{tikzpicture}[thick]
\fill[red] (1,0) circle (.07 cm);
\fill[red] (4,0) circle (.07 cm);
\draw[blue] (1 ,0) -- (4 ,0);
\node at (1,-0.4) {$D_\infty$};
\node at (2.5,-0.4) {$D_2$};
\node at (4,-0.4) {$D_\infty$};
\end{tikzpicture}  
\end{center}
\caption{Orbit space for a quadrupole, including isotropy groups.}
\end{figure}
Its boundary $\partial\Omega_2$ consists of two isolated 'singularities', namely the points $z=\pm 1$. These endpoints correspond to the maximal isotropy group $D_\infty$ (uniaxial symmetry). Points in the interior correspond to quadrupole tensors with generic isotropy group $D_2$ (biaxial symmetry). In a Landau-de Gennes potential one must include terms up to and including order six if symmetry breaking is to reach the $D_2$ phase (see \cite{TT}, \cite{VLF}, \cite{AL}).

\subsection{Three Vectors}
An example of a physical system which involves three vector order parameters is provided by electrons on a honeycomb or triangular lattice, doped to the saddle point of the band structure \cite{NCC}. These authors provide a sixth-order Ginzburg-Landau theory (see their Eqn. (8) and the subsequent discussion).\\ 

It is well-known \cite{We} that the minimal integrity basis of three vectors, $\bf a$, $\bf b$, $\bf c$, consists of the seven invariants
\begin{equation}
{\bf a}^2\ ,\quad {\bf b}^2 \ ,\quad {\bf c}^2 \ ,\quad 
{\bf a}\cdot {\bf b} \ ,\quad {\bf a}\cdot {\bf c} \ ,\quad 
{\bf b}\cdot {\bf c} \quad ,\quad 
[{\bf a},{\bf b},{\bf c}]\equiv\bf a\cdot(\bf b\times \bf c)
\end{equation}
The six scalars are the squared magnitudes of the vectors and the inner products between the vectors. The pseudo-scalar $[\bf a ,\bf b ,\bf c ]$ is the (signed) volume of the parallelepiped spanned by the vectors. There is a single polynomial relation among these invariants, namely
\begin{equation}\label{Sabc}
[{\bf a},{\bf b},{\bf c}]^2 = 
 {\bf a}^2 {\bf b}^2 {\bf c}^2 
-{\bf a}^2 ({\bf b}\cdot {\bf c})^2
-{\bf b}^2 ({\bf c}\cdot {\bf a})^2
-{\bf c}^2 ({\bf a}\cdot {\bf b})^2
+ 2({\bf a}\cdot{\bf b}) ({\bf b}\cdot{\bf c}) ({\bf c}\cdot{\bf a})
\end{equation}
This is non-negative and vanishes if-and-only-if the three vectors are coplanar.\\

We take the cosines of the angles between the normalized vectors as coordinates of the orbit space $\Omega_{111}$ and define
\begin{equation}\label{xyz}
x = {\bf b}\cdot {\bf c}\equiv \cos\alpha \quad ,\quad 
y = {\bf c}\cdot {\bf a}\equiv \cos\beta  \quad ,\quad
z = {\bf a}\cdot {\bf b}\equiv \cos\gamma 
\end{equation}
If we also define $p = [{\bf a},{\bf b},{\bf c}]$ then the polynomial relation (\ref{Sabc}) reads
\begin{equation}\label{pxyz}
p^2 = 1 - x^2 - y^2 - z^2 + 2xyz
\end{equation}
Substitution of (\ref{xyz}) shows this to be the familiar expression for the square of the volume of a parallelepiped with unit-length edges\footnote{Hence, any $SO(3)$-invariant polynomial function of three vectors must take the form $f(x,y,z)+g(x,y,z)p$ where $f$ and $g$ are polynomials in $x,y,z$. If analytical functions are allowed then one may eliminate $p$ as the square root of (\ref{pxyz}) but one must then still keep track of the sign of $p$.}. We thus find that the orbit space of three vectors is given by the inequality
\begin{equation}\label{Orbit3Vector}
x^2 + y^2 + z^2 - 2xyz \,\leq\, 1 \quad , \quad x,y,z\in [-1,1]
\end{equation}
We remark that this can be viewed as a one-parameter family of elliptic disks, stacked along any coordinate axis. Indeed, choosing for definiteness the $z$-axis as the stacking direction, we can write the orbit space inequality (\ref{Orbit3Vector}) as 
\begin{equation}
0\,\leq\,\frac{(x+y)^2}{1+z}\, +\,\frac{(x-y)^2}{1-z}\,\leq\, 2 
\quad ,\quad z\in [-1,1]
\end{equation}
This represents a set of ellipses with axes oriented at $45^{\rm o}$ to the $x$- and $y$-axes with semi-major and minor axes given by $\sqrt{1\pm z}$. The boundary surface is determined by
\begin{equation}\label{S3vector}
S(x,y,z)\,\equiv\, x^2 + y^2 + z^2 - 2xyz  - 1 = 0
\end{equation}
Allowing $x, y, z$ their full range, we recognize this as Cayley's famous nodal cubic surface\footnote{Not to be confused with Cayley's {\it ruled} cubic surface.} \cite{Ca}, the unique cubic surface with four isolated $A_1$ conic singularities \cite{H}. Hence, we have shown that the boundary of the orbit space for three vectors is given by the central part of Cayley's nodal cubic surface\footnote{See also R. Ferr\'eol at http://www.mathcurve.com/surfaces.gb/cayley/cayley.shtml .}.\\

The four nodes can be located by solving $S = {\bf\nabla}S = 0$ and are found to lie at the corners of a regular tetrahedron, namely 
\begin{equation}
(1,1,1),\ (1,-1,-1),\ (-1,\,1,-1),\ (-1,-1,\,1)
\end{equation}
The six straight edges connecting the cusps are also part of the boundary of the orbit space which has the shape of an inflated tetrahedron. See Figure 2.\\

\begin{figure}[H]
\centering
\includegraphics[width=0.6 \textwidth]{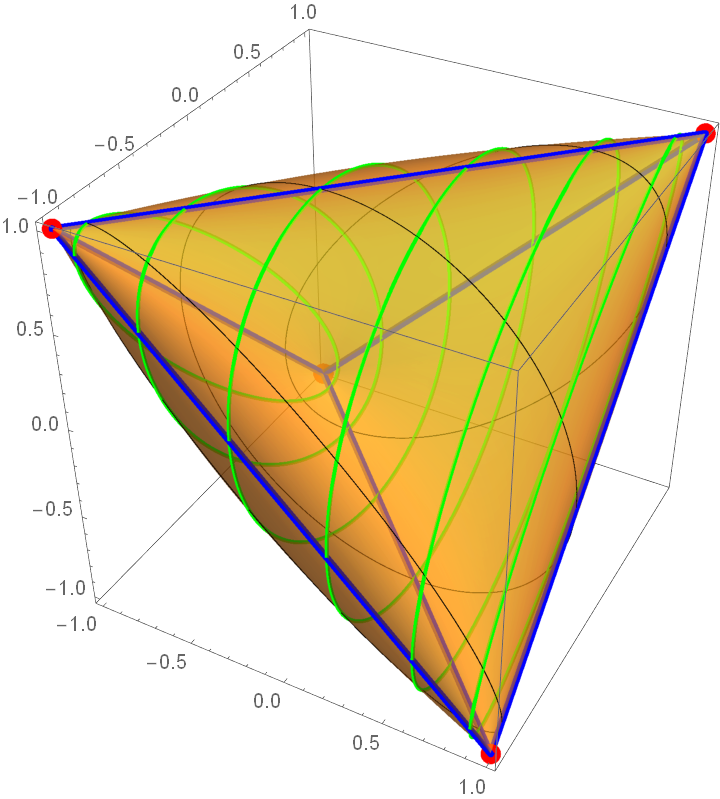}
\caption{Orbit space for three vectors with Cayley's nodal cubic surface as boundary.}
\end{figure}

As Eqn. (\ref{S3vector}) is quadratic in each variable, one easily solves it for $z$ to find
\begin{equation}  
z(x,y) \equiv xy \pm \sqrt{(1-x^2)(1-y^2)}
\end{equation}
This suggests the following, simple, trigonometric parametrization of the boundary 
\begin{equation}\label{cos-surface}
{\bf r}(\alpha, \beta) =(\cos\alpha,\cos\beta,\cos(\alpha+\beta))
\quad ,\quad \alpha\in [0,2\pi]\ ,\ \beta\in [0,\pi]
\end{equation}
This also follows from (\ref{xyz}) and coplanarity of the vectors, i.e. $\alpha +\beta +\gamma = 2\pi$, hence $z=\cos(\alpha+\beta)$. Lacking such geometrical insight, a systematic approach would start by choosing axes so that
\begin{equation}
{\bf a} = {\bf e}_1\sin\beta + {\bf e}_3\cos\beta \ ,\quad
{\bf b} =({\bf e}_1\sin\phi  + {\bf e}_2\cos\phi )\sin\alpha + 
              {\bf e}_3\cos\alpha  \ ,\quad
{\bf c} = {\bf e}_3  
\end{equation}
where $\phi$ is the azimuthal angle. This tells us that for any point in the orbit space
\begin{equation}\label{fac-abc}
z= \cos\alpha\,\cos\beta +\sin\alpha\,\sin\beta\,\cos\phi 
\quad ,\quad  p =\sin\alpha\,\sin\beta\,\sin\phi 
\end{equation}
A point lies on the boundary if-and-only-if $p=0$ which requires\footnote{Other choices, e.g. $\alpha=0$ or $\pi$, produce only part of the boundary surface.} $\phi=0$ or $\pi$. The choice $\phi=\pi$ recovers the parametrization (\ref{cos-surface}). The expression $p = \sin\alpha\,\sin\beta\,\sin\phi$ can be viewed as a factorization of the cubic equation (\ref{S3vector}) for the boundary. A similar approach wil be followed below for the orbit space of a vector plus quadrupole and also for an octupole.\\

\subsection{Vector $\oplus$ Quadrupole}
Liquid crystals are described by means of an $\ell = 2$ tensor as order parameter but, to account for permanent dipole moments or flexoelectrical effects, a secondary vector order parameter is introduced \cite{LT}. A picture of the corresponding orbit space seems to have appeared first in section 5A of \cite{K}, entitled $SU(3)$ adjoint + vector representations\footnote{The orbit space for $\bf 3\oplus 8$ of $SU(3)$ and that for $\bf 3 \oplus 5$ of $SO(3)$ are identical as they formally have the same invariants and the same syzygy.}, but no equation or parametrization was given there. Such expressions were given in \cite{LT} but the focus there was on the possible phases, not on the geometry of the orbit space. We will find here a new, simpler, trigonometric parametrization of this orbit space and identify its boundary surface as a particular quartic ruled surface.\\

A vector $\bf a$ plus a quadrupole $A$ possess the following invariants 
\begin{equation}
{\bf a}^2 \ ,\quad {\rm tr} A^2 \ ,\quad {\bf a}\cdot A{\bf a} \ ,\quad
{\bf a}\cdot A^2{\bf a}\ ,\quad{\rm tr} A^3\ ,\quad [{\bf a}, A{\bf a},A^2{\bf a}] 
\end{equation}
The unique polynomial relation among these invariants is obtained from the relation (\ref{Sabc}) for three vectors by the substitutions ${\bf b}\rightarrow A{\bf a}$, ${\bf c}\rightarrow A^2{\bf a}$ and repeated application of the Hamilton-Cayley identity (\ref{HC-id}). We normalize $\bf a$ and $A$ and define orbit space coordinates and pseudo-scalar by\footnote{The relation of our coordinates to those of \cite{K} is: $x =\sqrt{\frac32}\,\beta_1$, $y = 3\beta_2 - 1$, $z=\sqrt{6}\,\alpha_3$.} 
\begin{equation}
x = \sqrt{\frac32}\ {\bf a} \cdot A {\bf a} \ ,\quad 
y = 3\, {\bf a} \cdot A^2 {\bf a} - 1     \ ,\quad
z = \sqrt{6}\,{\rm tr} A^3                \ ,\quad 
p = [ {\bf a}, A{\bf a}, A^2 {\bf a} ]
\end{equation}
Hence, the orbit space for a vector plus a quadrupole is given by the inequality
\begin{equation}\label{ineq2}
p^2 = 1- 3x^2 - 3y^2 - z^2 + 6x^2 y - 2y^3 + 6xyz - 4x^3 z \,\geq\, 0 
\, ,\ x,y,z\in [-1,1]
\end{equation}
We note that this inequality can be written in the compact form
\begin{equation}
(2x^3 -3xy+z)^2 \leq (2x^2 -y -1)^2 (x^2 -2y +1)
\end{equation}
Hence, $y =2x^2 - 1$ implies $z =(4x^2 - 3)x$. Similarly, 
$y =\frac12 (x^2 + 1)$ implies $z =\frac12 (3 - x^2)x$  (below, these will be shown to correspond to the double curve and the striction curve, respectively).\\

Coplanarity of ${\bf a}$, $A{\bf a}$ and $A^2{\bf a}$ requires these vectors to lie in a plane spanned by a pair of eigenvectors of $A$. To prove this, consider first a diagonal matrix $M = {\rm diag}[\mu_1 ,\mu_2, \mu_3]$ which acts on a vector $\bf v$. One verifies easily that the pseudo-scalar built out of $\bf v$ and its $M$-iterates satisfies
\begin{equation}
[{\bf v}, M{\bf v}, M^2{\bf v}] = v_1 v_2 v_3\,\Delta(\mu_1,\mu_2,\mu_3) 
\ , \quad \Delta(\mu_1,\mu_2,\mu_3) = \prod_{i<j} (\mu_j - \mu_i)
\end{equation}
where the Vandermonde determinant of $M$ makes its appearance. Hence, the pseudo-scalar vanishes if $\Delta$ or (at least) one component of $\bf v$ vanishes. We now apply this to our quadrupole tensor $A$ in its eigenbasis (\ref{Adiag}) and we set ${\bf v} = {\bf a}$ where
\begin{equation}
{\bf a} = ({\bf e}_1 \cos\phi + {\bf e}_2 \sin\phi )\sin\theta + 
{\bf e}_3\cos\theta  
\end{equation}
This yields a factorization, similar to that in (\ref{fac-abc}), of $p=[{\bf a}, A{\bf a},A^2{\bf a}]$, namely
\begin{equation}
p\ \propto\ \sin\theta\,\sin 2\theta\,\sin 2\phi\,\sin 3\alpha 
\end{equation}
Again, a point lies on the boundary surface if-and-only-if $p$ vanishes. The optimal choice turns out to be $\theta = \pi/2$ which places ${\bf a}$ in the 12-plane and yields as parametrization
\begin{equation}\label{RuledSurface1}
{\bf r}(\alpha,\beta) = 
( -\tfrac12 \cos \alpha,\, -\tfrac12\cos 2\alpha,\,\cos 3\alpha )
 +\tfrac12\sqrt{3}\,(-\sin\alpha,\,\sin 2\alpha,\,0)\cos\beta
\end{equation}
with $\alpha,\,\beta\equiv 2\phi \in [0,\pi]$.
This parametrization proves that this is a ruled surface\footnote{A ruled surface is swept out by a line in space, hence it has a local parametrization of the form
${\bf r}(t,u) = {\bf b}(t) + u {\bf d}(t)$
where ${\bf b}(t)$ and ${\bf d}(t)$ are known as the base and director curves, respectively.}. 
The so-called standard parametrization of this surface is
\begin{equation}\label{RuledSurface2}
{\bf r}(t,u) =\left(t,\,\tfrac12 (1+t^2),\,\tfrac12 (3-t^2)t \right) 
+ \frac{u}{\sqrt{1+t^2}} (1, t, 0)
\end{equation}
See Figure 3. Note that this orbit space is a stack of triangles with the corners of each triangle lying on the twisted cubic curve. 

\begin{figure}[h]
\centering
\includegraphics[width=0.6 \textwidth]{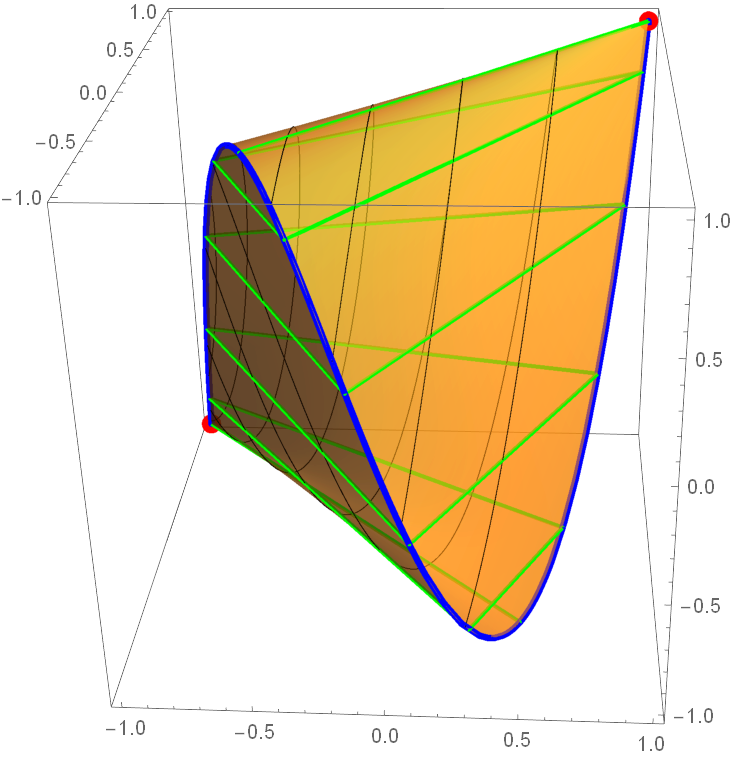}
\caption{The orbit space for a vector plus quadrupole with a quartic rule surface as (translucent) boundary.}
\end{figure}

The double curve, i.e the curve on which two generators meet, is the twisted cubic
\begin{equation}
y = 2x^2 - 1 \quad , \quad z = 4x^3 - 3x \quad ,\quad x\in [-1,1]
\end{equation}
An elegant parametrization of the double curve is thus given by
\begin{equation}
x = \cos\alpha \ ,\quad  y =  \cos 2\alpha \ , \quad 
z =  \cos 3\alpha \quad ,\quad \alpha\in [0,\frac{\pi}3]
\end{equation}
The striction curve is readily found to be parametrized by the base vector, i.e. the first vector on the right hand side of either (\ref{RuledSurface1}) or (\ref{RuledSurface2}).\\

%Isotropy groups: See Notizen 02-12-2016 and 04-03-2018.\\
The strata of the orbit space and the associated isotropy groups can be decsribed as follows: 
When $\bf a$ lies in a plane spanned by precisely two eigenvectors of a generic, i.e. non-degenerate, $A$ then the configuration $\{{\bf a}, A\}$ gets mapped to a point on the boundary surface, not on the double curve; the isotropy group is $C_2$.  When $\bf a$ is proportional to a single eigenvector of generic $A$ then $\{{\bf a}, A\}$ gets mapped to a point on the double curve;  the isotropy group is $C_2 \times C_2$. For degenerate $A$ and $\bf a$ proportional to the non-degenerate eigenvector, $\{{\bf a}, A\}$ gets mapped to the endpoints of the double curve; the isotropy group is $D_\infty$.\\ 

Quartic ruled surfaces were first classified from a projective geometric point of view by Cremona, Cayley, and Steiner and studied recently again \cite{PPT}. Based on the fact that the double curve is a twisted cubic curve, we can identify our ruled surface as the first species in Cremona's list which had a total twelve species\footnote{This is case 14 in \cite{PPT} where it is stated that there are six distinct subclasses for this case.}. We further note that there exist exactly two tangents to the double curve which are at the same time generators of the extended ruled surface.\\  

It may appear that the boundary surface is the convex hull of a segment of the twisted cubic curve but this is not the case. A convex hull is a ruled surface so it has non-positive Gaussian curvature $K$ at all points, but as it is convex its curvature cannot be negative so $K$ has to vanish. Hence, any convex hull is a developable surface. We found that our orbit space is ruled but not developable, hence it cannot be the convex hull of the twisted cubic curve segment we have been dealing with\footnote{In fact, the convex hull of a finite segment of a twisted cubic curve consists of the union of two oblique cones whose vertices are the two endpoints of the curve, the straight line segment between those points being the place where the cones meet. A parametrization of the two components of this convex hull is given by $(1-u)\, (\cos\theta,\cos 2\theta,\cos 3\theta)  + u\, (\pm 1, 1, \pm 1)$ with $\theta\in [0,\pi]$, $u\in [0,1]$.}.
%  or  (t,t^2,t^3) + u (t\pm 1,t^2-1,t^3\pm 1)  
% Notizen 26-12-2016  and 19-03-2017

\section{Octupole Tensor}
We will now investigate the geometry of the orbit space of an octupole, a traceless symmetric third-rank tensor. Physical systems with an octupole tensor as order parameter include liquid crystals with bent, banana-shaped, molecules \cite{LuR} and Kagome lattices \cite{Zh}. Earlier work includes that of \cite{O} where $SO(3)$ tensors were decomposed into their isotropy classes, a method going back to \cite{Sir}, but the geometry of the orbit spaces was not studied there; see also \cite{BN}. For the case $\ell = 3$, \cite{O} correctly listed $SO(2)$, $D_3$, and $T$ as the possible maximal isotropy groups but in the decomposition of an $\ell = 3$ tensor a term is missing which we identify as the orientation angle of the '3-star', i.e. the $D_3$ component\footnote{This angle appears only in the degree ten invariant which plays no r\^ole in the Higgs potential.}. In \cite{K}, only coregular representations were studied but it is also stated there that "The $SO_3$ seven-dimensional representation has four maximal little groups. If its orbit space is built out of polynomials of degree 2, 4, 6, and 10, then it is a warped tetrahedron". As we will show, this is not correct.\\

The possible isotropy, or stability, subgroups for the $\ell = 3$ irreps of $SO(3)$ are in Sch\"onflies notation given by 
$C_\infty,\, T,\, D_3,\, C_3,\, C_2$ and $C_1$ (see \cite{LS} which includes a detailed discussion of past errors). For the group $O(3)$ one must distinguish the irreps $\ell^\pm$. For $3^-$ the possible isotropy subgroups are then $C_{\infty v},\, T_d,\, D_{3h},\, C_{3v},\, C_{2v},\, C_s$ and $C_1$. For $3^+$ the list is $C_{\infty h},\, T_h,\, D_{3d},\, C_{3i},\, C_{2h}$ and $C_i$. We will focus here on the $\ell =3$ irreps of $SO(3)$. Given that there are then three maximal isotropy subgroups, namely $C_\infty,\, T$, and $D_3$, and that, as we will see, there are four relevant invariants, all even in the tensor, we expect the orbit space to be a three-dimensional body with three cusps.\\

We call the $\ell = 3$ tensor $B_{ijk}$, assume it to be non-vanishing, and define the associated quadrupole tensor $A$ and vector ${\bf a}$ by
\begin{equation}
A_{ij} = B_{ikl} B_{jkl} - \frac13 B_{klm} B_{klm}\,\delta_{ij} \quad , \quad  a_i = B_{ijk} A_{jk} 
\end{equation}
This quadrupole tensor has of course a canonical, diagonal, form. No such form would seem to exist for the $\ell = 3$ tensor itself. However, we note that $A$ is the unique quadrupole tensor associated with $B$. Hence, we choose to define the canonical form of an $\ell = 3$ tensor to be its appearance in the eigenbasis of its associated quadrupole tensor\footnote{A binary sextic has five covariants of order two. Those of degree 2 and 4 correspond to the tensors $A$ and $A^2-{\rm trace}$, respectively, which have a common eigenbasis. Those of degree 5,7, and 9, correspond to axial tensors. An $\ell > 3$ tensor has many covariants of order two, hence no unique canonical form.}.\\

Classical invariant theory \cite{C} informs us that a binary sextic, the $SL(2,C)$ version of our $\ell = 3$ tensor, possesses a Hilbert basis consisting of five invariants $I_n$ of degrees $n$ = 2, 4, 6, 10 and 15; cf. Theorem 2.1 of \cite{OA}. Our explicit Hilbert basis is then given by (see also Figure 4)
\begin{equation}\label{MIBoctupole}
I_2 = B_{ijk} B_{ijk} \ , \  I_4 = {\rm tr} A^2 \ ,\  I_6 = {\rm tr} A^3 \ ,\ 
I_{10} = {\bf a}\cdot A^2{\bf a}  \ , \  I_{15} = [{\bf a}, A{\bf a},A^2{\bf a}] 
\end{equation}

\begin{figure}
\begin{center}
\begin{tikzpicture}[thick] 
\draw[thick] (0.7, 0) circle (0.7 cm); 
\draw (0 ,0) -- (1.4 ,0); 
\node at (0.7,-1.4) {$I_2$}; 
\draw[thick] (4,0) circle (1 cm); 
\draw (3.133 ,0.5) -- (4.866 ,0.5); 
\draw (3.133 ,-0.5) -- (4.866 ,-0.5); 
\node at (4,-1.4) {$I_4$}; 
\draw[thick] (7,0) circle (1 cm); 
\draw (6.39 ,0.79) -- (7.61 ,0.79); 
\draw (7.99, 0.13) -- (7.38,-0.92); 
\draw (6.617 ,-0.924) -- (6.009 ,0.131); 
\node at (7,-1.4) {$I_6$}; 
\draw[thick] (9.7,0) circle (0.7 cm); 
\draw (9.7,0.7) -- (9.7,-0.7); 
\draw (10.4,0) -- (10.9 ,0); 
\draw[thick] (11.2,0) circle (0.3 cm); 
\draw (11.5,0) -- (12.0,0); 
\draw[thick] (12.3,0) circle (0.3 cm); 
\draw (12.6,0) -- (13.1,0); 
\draw[thick] (13.8,0) circle (0.7 cm); 
\draw (13.8,0.7) -- (13.8 ,-0.7); 
\node at (11.8,-1.4) {$I_{10}$}; 
\end{tikzpicture}
\end{center}
\caption{Graphical representation of the elements of our Hilbert basis (\ref{MIBoctupole}). Vertices correspond to $B$-tensors. The subgraph -o-  represents the {\it traceless} tensor $A$.}
\end{figure}
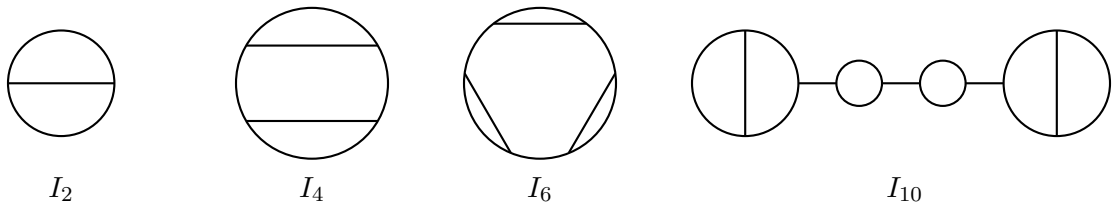
In particular, any invariant of degree eight is reducible and can be expressed as a polynomial in $I_2$, $I_4$ and $I_6$ as e.g. in ${\bf a}\cdot A {\bf a} = \frac23 (I_2 I_6 + {I_4}^2)$. Instead of $I_4$, one may use the tetrahedral invariant $J_4 = B_{ijk} B_{ilm} B_{jln} B_{kmn}$, the two being related by $I_4 + J_4 = \frac16{I_2}^2$. There is even more choice for higher degree invariants (see the Appendix). \\ 

The pseudo-scalar $I_{15}$ has the same structure as in the three vector case. Hence, the syzygy is again obtained by taking the square of the pseudo-scalar $I_{15}$ and expressing it in terms of the scalar invariants. In the process, repeated use is made of the Cayley-Hamilton theorem for $A$. This yields a single polynomial relation of degree thirty in $B$, first found by Clebsch \cite{C} using so-called transvectants 
({\it\" Uberschiebungen}). Although such notation is very effective we will continue to use tensor notation as it is more familiar to physicists. Before writing Clebsch's syzygy, we set $I_2 = 1$ and define our intial orbit space coordinates by
\begin{equation}\label{original-xyz}
x = 6 I_4 \ ,\quad y = 36 I_6  \ ,\quad  z = 162 I_{10}
\end{equation}
with normalizations chosen such that $x\in [0,1]$ and $y\in [-1,1]$. 
% This y = 2 old-y - x$ where old-y = 9K6 and  old-z = 2 this z. 
The $z$-coordinate can be redefined in non-trivial ways. Indeed, the most general change of the invariants in the Hilbert basis (\ref{MIBoctupole}) which preserves homogeneity takes the form
\begin{eqnarray}
I_4\rightarrow a_0 I_4 + a_1 {I_2}^2 \ , \quad
I_6\rightarrow b_0 I_6 + b_1 I_2 I_4 + b_2 {I_2}^3 \ , \quad & & \nonumber\\
I_{10}\rightarrow c_0 I_{10} + c_1 I_2 {I_4}^2 
+ c_2 I_4 I_6 + c_3 {I_2}^3 I_4 + c_4 {I_2}^2 I_6 + c_5 {I_2}^5 & &
\end{eqnarray}
with $a_0,\,b_0,\,c_0\neq 0$. Setting $I_2 = 1$, the most general redefinition of the orbit space coordinates is then seen to take the form (with $a_{ii} \neq 0$)
\begin{eqnarray}\label{redef}
x\rightarrow a_{11} x + v_1 \ , \quad
y\rightarrow a_{12} x + a_{22} y + v_2\ , \quad & & \nonumber\\
z\rightarrow a_{31} x + a_{32} y + a_{33} z + v_3 + k_1 x^2 + k_2 xy  \quad & &
\end{eqnarray}
This shows that we may change the coordinate $z$ by terms non-linear in $x$ and $y$. We remark that for $k_1 = k_2 = 0$ this is a lower triangular affine transformation ${\bf r} \rightarrow A{\bf r} + {\bf v}$. The triangularity of the affine matrix $A$ can be traced to the original orbit space being a weighted projective space. We now use the freedom to redefine the orbit space coordinates as in (\ref{redef}) to offer three distinctive pictures of the orbit space of an octupole.\\

\subsection{Orbit Space of an Octupole: A First View}
In terms of the initial coordinates (\ref{original-xyz}), the Clebsch syzygy reads
\begin{eqnarray}\label{Clebsch1}
16 z^3 - 12 x (x + y) z^2 - 12(x^2 + y) (x^3 + 2xy + y^2) z 
+ 9 x^7 + (24 + 13x) x^5 y & & \nonumber\\
+(16 +42x) x^3 y^2 +(36 +6x) x^2 y^3 +(4 +9x) y^4 + y^5\ \leq\ 0 & &
\end{eqnarray}
This polynomial has total degree seven but it is only cubic in $z$ and it has a simple discriminant with respect to $z$, namely
\begin{equation}
\Delta = 2^8 3^3\,(x^3 -y^2)^3 \,\Big(2(1-x)^3 - (y -2 +3x)^2\Big)^2 
\end{equation}
Hence, roots in $z$ coincide on two semi-cubical parabolas in the $xy$-plane, namely $y^2 = x^3$ and $(y -2 +3x)^2 = 2(1-x)^3$ (this coninues to hold for the other pictures of the orbit space offered below); see Fig. 5.\\

\begin{figure}[H]
\centering
\includegraphics[width=0.5 \textwidth]{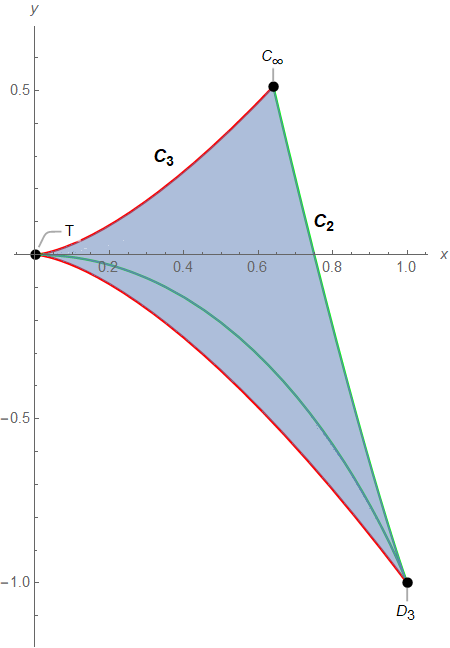}
\caption{Projection of the orbit space for an octupole on the $xy$-plane. The boundary curves are branches of two semi-cubical parabolas. The isotropy groups are included.}
\end{figure}
A rational parametrization of the boundary surface is given by
\begin{eqnarray}\label{ratpar1}
x = t^2 + 2tu\ ,\ 
y = t^3 + 3t^2 u\ , \  
z = t^4 +t^5 + (4t^3 +5t^4) u + \frac12 (6 +13t) t^2 u^2
& \quad &\\
t \in [-\frac65,\,2]\, , \  
u \in [u_-,\,u_+]\  ,\   u_\pm = \frac1{64t} 
\,\Big(12-4t-25t^2 \pm (2-t)\sqrt{3(2-t)(6+5t)}\Big)
& &\nonumber
\end{eqnarray}
The indicated limits for $t$ and $u$ remain the same in the sequel, hence will not be repeated. There are two cuspidal edges which we label with the corresponding isotropy group:  
\begin{eqnarray}
&C_3\ :& {\bf r}_1 (t) = (t^2,\,t^3,\,t^4 + t^5)
\qquad ,\qquad t\in [-1,\,\frac45]  \nonumber\\
&C_2\ :& {\bf r}_2 (s) = 
\Big(1-s^2,\,(1+s)^2(2s-1),\,\frac12 s^2 (1+s)^3\Big)
\ ,\ s\in [-1,\,\frac35]
\end{eqnarray}
The cusps, also labeled by their isotropy group, are given by
\begin{eqnarray}\label{cusps1}
&T\ :& {\bf r}_1(0) = {\bf r}_2(-1) = (0,\,0,\,0)\nonumber\\  
&D_3:& {\bf r}_1(-1) = {\bf r}_2(0) = (1, -1,\,0) \\  
&C_\infty :& {\bf r}_1(\frac45) = {\bf r}_2(\frac35) = 
(\frac{16}{25},\,\frac{64}{125},\,\frac{2304}{3125}) \nonumber
\end{eqnarray}
Our first picture of the three-dimensional orbit space of an octupole is then

\begin{figure}[H]
\centering
\includegraphics[width=0.6 \textwidth]{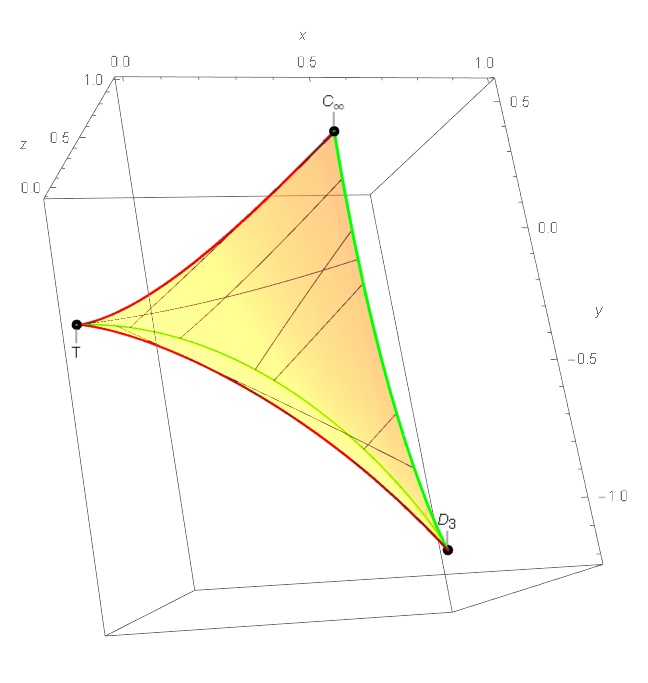}
\caption{A first, translucent, view of the orbit space for an octupole as in Eqns. (\ref{Clebsch1}-\ref{cusps1}).}
\end{figure}

As expected on general grounds, this orbit space has three cusps\footnote{We call the point at which the isotropy is $C_\infty$ a cusp but, upon extending our view in section 4 to the moduli space for a binary sextic, only the points at which the isotropy group is $T$ or $D_3$ are cusps.}. 

\subsection{Orbit Space of an Octupole with a Planar Cuspidal Edge}
Define $Z = 2x (x + y) - 2z$. In terms of the  coordinates $x, y, Z$, Clebsch's syzygy becomes
\begin{equation}\label{Clebsch2}
(4-3x + y) (x^3 - y^2)^2 \,\leq\, 
\Big( 2Z^2 -9 x (x + y) Z 
- 6(x^3- y^2) (x^2 - 2x - y)\Big) Z
\end{equation}
Note the compactness of this expression as compared to that in (\ref {Clebsch1}). This form of the syzygy tells us e.g. that $Z=0$ implies $y^2 = x^3$ (the alternative $y = 3x - 4$ lies, except for cusp $D_3$, outside the orbit space.). The rational parametrization (\ref{ratpar1}) of the boundary surface simplifies into
\begin{equation}\label{ratpar2}
x = t^2 + 2 t u   \quad , \quad 
y = t^3 + 3 t^2 u \quad , \quad 
Z = (2-t)\, t^2 u^2
\end{equation}
A trigonometric parametrization of the boundary surface is given by
\begin{equation}
x = \mu^2              \quad , \quad 
y = \mu^3 \cos 3\alpha \quad , \quad 
Z = \frac12\,\mu^4 (1 + \mu\cos\alpha) (1+2\cos 2\alpha)^2
\end{equation}
In these coordinates, the cuspidal edges become
\begin{eqnarray}
&C_3\ :& {\bf r}_1 (t) = (t^2,\,t^3,\,0) \nonumber\\
&C_2\ :& {\bf r}_2 (s) = 
\Big(1-s^2,\,(1+s)^2(2s-1),\, s^2 (1+s)^2 (3-5s)\Big)
\end{eqnarray}
Thus, the cuspidal edge ${\bf r}_1 (t)$ is a planar semi-cubical parabola, lying in the $xy$-plane, as do the cusps:
\begin{eqnarray}\label{cusps2}
&T\ :&\  {\bf r}_1(0) = {\bf r}_2(-1) = (0,\,0,\,0)  \nonumber\\
&D_3:\ & {\bf r}_1(-1) = {\bf r}_2(0) = (1, -1,\,0)\ \\
&C_\infty :\ &  {\bf r}_1(\frac45) = {\bf r}_2(\frac35) = (\frac{16}{25},\,\frac{64}{125},\,0) \nonumber
\end{eqnarray}
\pagebreak 

A second picture of the orbit space of an octupole is hence 

\begin{figure}[H]
\centering
\includegraphics[width=0.6 \textwidth]{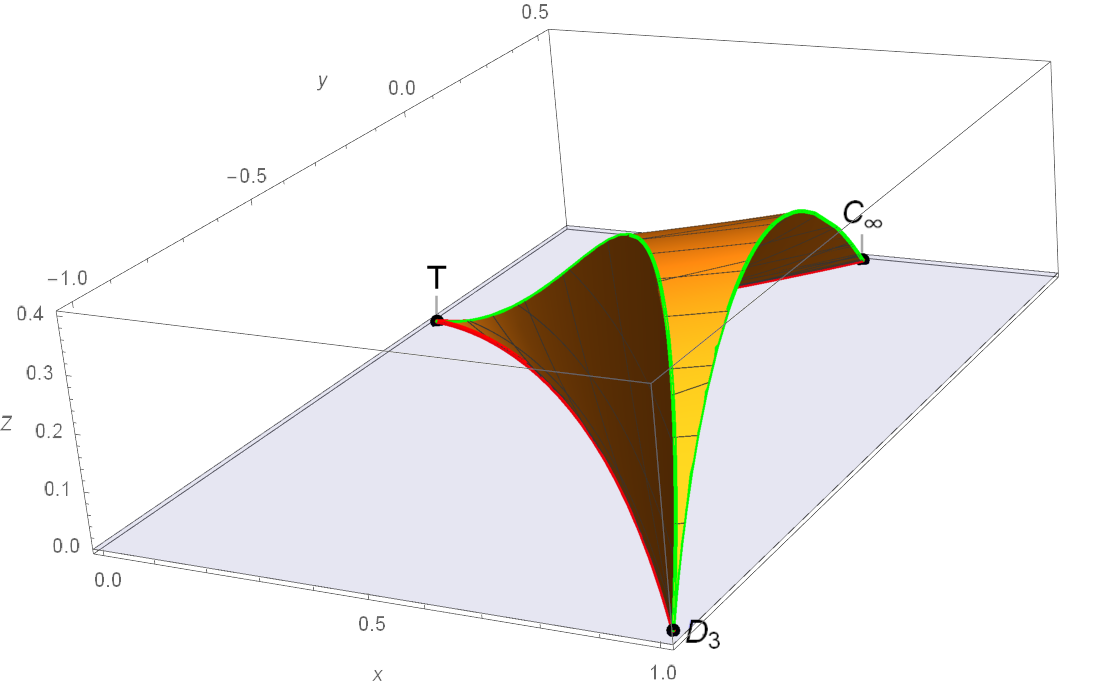}
\caption{The orbit space for an octupole with a planar cuspidal edge as in Eqns. (\ref{Clebsch2}-\ref{cusps2}).}
\end{figure}

The orbit space now looks somewhat like a conical roof. Its boundary surface is not ruled, let alone developable.\\

\subsection{Orbit Space of an Octupole as a Tangent Developable}
Define $w = \frac16 (3x - y) x - Z$. In terms of the coordinates $x, y, w$, Clebsch's syzygy reads
\begin{eqnarray}
w^3 + (3x + 5y) x w^2  
+(\frac94 x^4 - 3 x^5 - \frac12 x^3 y  - 6 x y^2 +\frac{55}{12} x^2 y^2 - 3 y^3)\, w \ & & \nonumber\\
+ \frac16\Big(8x^4 -6x^3 +3x^2 y -\frac{20}9 x^3 y  
+ 12(1 - x) y^2 + 3y^3\Big)\,y^2 &\leq\ 0 &\qquad\qquad
\end{eqnarray}
The total degree has been reduced to six but, more importantly, the boundary surface has become a tangent developable\footnote{A tangent developable is a ruled surface swept out by the tangent to a space curve, known as the regression curve. Hence, it is a special ruled surface ${\bf r}(t,u) = {\bf b}(t) + u {\bf d}(t)$ for which the director curve ${\bf d}(t)$ equals the derivative of the base curve ${\bf b}(t)$.}. Indeed, the surface now has rational parametrization
\begin{equation}\label{TD}
{\bf r}_1 (t) + u\,{\bf r}_1' (t) \ ,\quad 
{\bf r}_1 (t)= (t^2,\,t^3,\,\frac12 t^4 -\frac16 t^5) 
\end{equation}
which manifestly shows it to be a tangent developable with ${\bf r}_1 (t)$ as its regression curve. We have verfied that the Monge-Amp\`ere equation, $w_{xx}w_{yy} = w_{xy}^2$, is satisfied, offering independent confirmation of this fact. A calculation of the Gaussian curvature shows that it indeed vanishes at all points of the surface away from the cusps and cuspidal edges. The coordinates in which the surface is developable are unique up to lower triangular affine transformations as in (\ref{redef}). In order to sweep out the entire surface, the tangent line to the regression curve has to follow part of the extension of this curve beyond the cusps; see (\ref{ratpar1}). The implicit equations for the (extended) regression curve are $y^2 = x^3,\ (6w - 3x^2)^2 = x^5$. The double curve is given by
\begin{equation}
{\bf r}_2 (s) = \Big(1 - s^2,\,
(s + 1)^2(2s - 1),\,
\frac23 (s + 1)^2(2s - 1)^2(2s + 1)\Big) 
\end{equation}
and a trigonometric parametrization of the boundary surface is given by
\begin{equation}
x = \mu^2\, ,\ 
y = \mu^3 \cos 3\alpha\, ,\  
w = -\frac23\,\mu^4\cos^2\alpha \, 
\Big( 6\cos 2\alpha +\mu (4\cos\alpha + 3\cos 3\alpha )\Big)
\end{equation}
The cusps are now at the following locations:
\begin{equation}
T: \ (0,0,0) \ ,\quad 
D_3 : \ (1,-1,\,\frac23) \ ,\quad 
C_\infty :\ (\frac{16}{25}, \,\frac{64}{125}, \,\frac{1408}{9375})
\end{equation}
The above equations for the orbit space of an octupole and the tangent developable nature of its boundary surface constitute our main results. We are not aware of a priori reasons why it should have been possible to choose coordinates which make the boundary surface a tangent developable (but see also our Conclusions).\\

\begin{figure}[H]
\centering
\includegraphics[width=0.6 \textwidth]{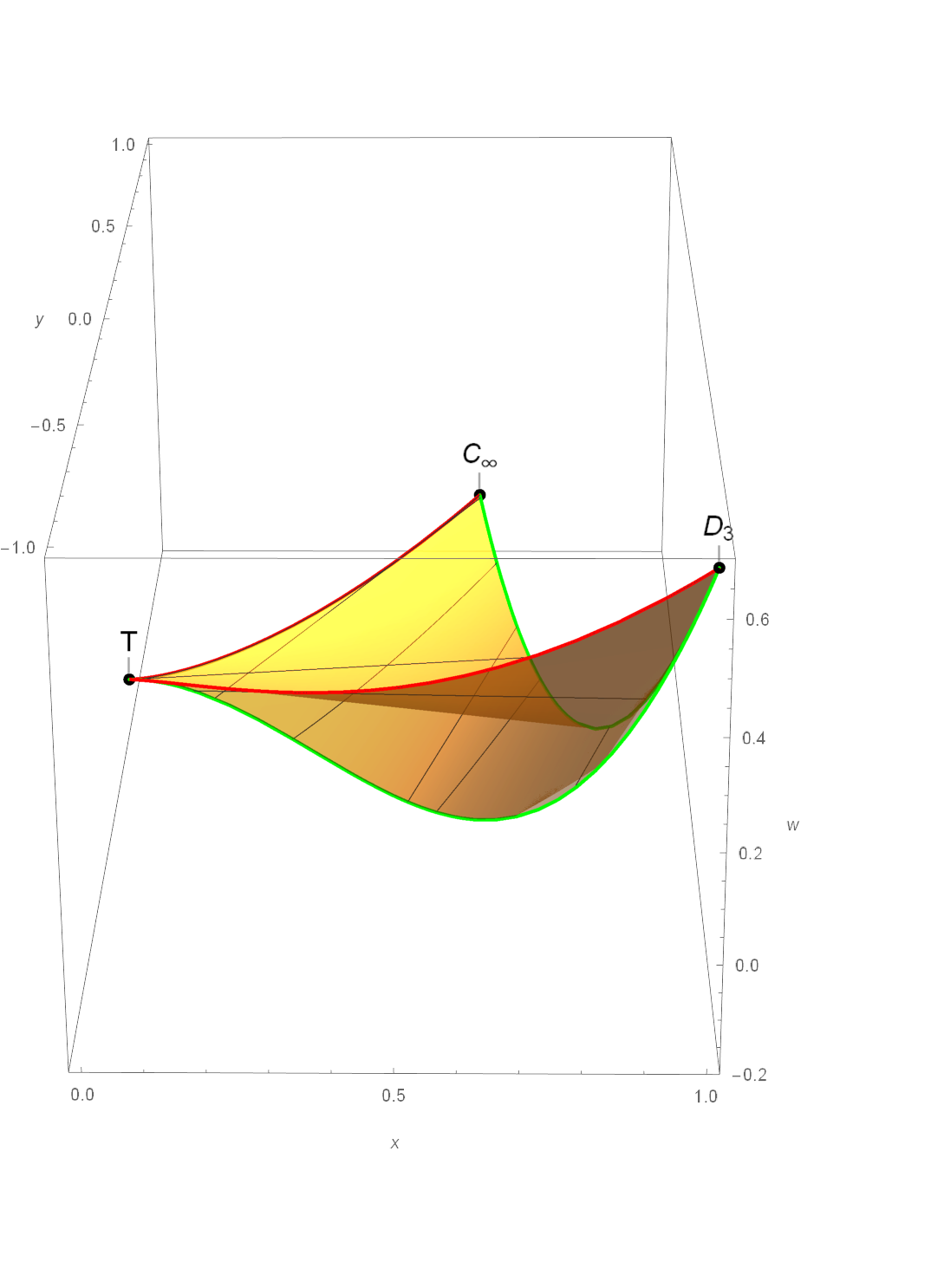}
\caption{The orbit space for an octupole in tangent developable form (translucent).}
\end{figure}

Being among the simplest of curved surfaces, tangent developables were studied long ago. In particular, those of degree six were classified from a projective geometric point of view by Chasles, Cayley, and others. We identify our developable surface with case 2 in the table on p. 286 of \cite{Ed}; see also p. 360 of \cite{Pa}. 
% N. 26-06-2018 and 16-03-2018

\subsection{Nature of the Singularities of the Extended Boundary Surface}
The number and nature of the cusps is preserved not only under polynomial transformations but also under diffeomorphisms. The kind of singularities which are possible for tangent developables have been studied by mathematicians. Izumiya \cite{Iz} reports that a tangent developable can only have cuspidal edges, cross-caps and swallowtails as singularities and Cleave \cite{Cl} has shown that the tangent developable of a space curve has a cross-cap at any point where the torsion $\tau$ vanishes with $\tau'\neq 0$. We use these insights here to classify the nature of the three point singularities of our orbit space\footnote{Strictly speaking, we should classify the singularities from a real perspective. Recently, programs were developed for this task \cite{MS} based on the software S{\tiny INGULAR} but only for low corank.}. To do so, we must first extend our surface by relaxing the constraints, set in (\ref{ratpar1}), on the parameters $t$ and $u$ in (\ref{TD}) (see also the next section).\\

We find the cusp where the isotropy group is $T$ to be a swallowtail, i.e. type $A_4$ in Arnol'd's classification. Indeed, $(x, y,2w) \approx (t^2,t^3,t^4)$ for small $t$ and it is known that the tangent developable to a space curve of type $(2,3,4)$ is a swallowtail. Also the cusp where the isotropy group is $D_3$ is a swallowtail as for small $s$ the double curve behaves as $(s^2,s^3,s^4)$ (this requires a lower triangular affine transformation of the coordinates). Finally, we find that at the point where isotropy is $C_\infty$ the torsion $\tau$ of the regression curve vanishes but $\tau' \neq 0$ there. Furthermore, near this point the regression curve is of type $(1,2,4)$. Hence, we see that the surface has a cross-cap at the $C_\infty$ point. Since a cross-cap is a M\"obius band, we conclude that the extended boundary surface is non-orientable.

\section{Relation to the Moduli Space of Genus Two Hyperelliptic Curves}
In order to find a Hilbert basis for our $\ell = 3$ tensor we appealed to the known basis of a binary sextic, a result of Classical Invariant Theory. This is based on a well-known link, called the Cartan map, between traceless symmetric rank-$\ell$ tensors of $SO(3,C)$ and complex binary forms of order $2\ell$ (see Appendix A of \cite{OKDD}). In order to represent real tensors of $SO(3)$, we restrict to $SU(2)$ binary forms, i.e. binary forms which satisfy 
\begin{equation}
b(\xi,\eta) = \sum_{k=0}^{2\ell} a_k \xi^k \eta^{2\ell -k} \quad ,\quad a_{2\ell -k} = (-1)^{\ell - k} {\bar a}_k
\end{equation} % ${\bar b}(-\eta,\xi)= (-1)^\ell b(\xi,\eta)$
Apart from this constraint then, we have actually been studying the moduli space of a binary sextic which in turn is of importance to the understanding of the moduli space of hyperelliptic curves of genus two. Also, what we have been calling the isotropy group of the octupole tensor is referred to by mathematicians as the automorphism group of the binary form. In general, there are more choices of automorphism group for the binary form then there are isotropy groups for the corresponding tensor. Still, for each isotropy group there is a definite automorphism group and we note here that the geometry of the orbit space for a rank-$\ell$ tensor of $SO(3)$ is (a real section of) the moduli space for a binary form of order $2\ell$. We will now describe this in more detail for a quartic and a sextic.\\

The invariants of a binary quartic describe the moduli space of elliptic curves of genus one. In that context, one assumes that the discriminant of the quartic is non-zero, ${I_2}^3 \neq 6{I_3}^2$, and one chooses the so-called absolute $j$-invariant, $j= {I_2}^3 /({I_2}^3-6 {I_3}^2)$, as coordinate of the moduli space. We instead assume $I_2 \neq 0$ and otherwise find the relation to our orbit coordinate $z$, Eqn. (\ref{z-coordinate}), to be 
\begin{equation}
j = \frac1{1-z^2} = \csc^2 3\alpha
\end{equation}
Hence, (a real section of) the moduli space of elliptic curves is the same as the orbit space for an $\ell = 2$ quadrupole tensor (Fig. 1).\\

Turning to the binary sextic, there is a study dating from 1887 by Oskar Bolza \cite{Bo} in which all possible automorphism groups are listed for this case\footnote{We disovered this paper only after completing our work on the orbit space of the octupole tensor.}. Bolza also gave the polynomial relations between the invariants of the sextic which select a particular automorphism group. These relations then define the strata of the moduli space of a binary sextic. Bolza used the Clebsch invariants $A,B,C$ and $D$ but these are identical to our invariants $I_2, I_4, I_6$ and $I_{10}$. It is then easy to translate Bolza's relations into ours and we obtained a complete match. Bolza labeled the six possible automorphism groups with Roman numerals I through VI. The correspondence with the strata of our orbit space and their associated isotropy groups is as follows:\\

VI: $I_2 \neq 0$, $I_4 =0,\ I_6 = 0,\ I_{10} = 0$. Cusp at $(0,0,0)$ with isotropy group $T$.\\

V\ : $I_2\neq 0$, ${I_2}^2 = 6I_4$, $I_2 I_4 + 6I_6 = 0$, $I_{10} = 0$. Cusp at $(1,-1,0)$, isotropy $D_3$.\\

IV: ${I_4}^3 = 6{I_6}^2$, $2I_4 (I_2 I_4 +6I_6) = 9I_{10}\neq 0$, $2I_2 I_4 \neq 15 I_6$.\\
Cuspidal curve $y^3 = x^2$, $z = 2x (x + y)$, $4x \neq 5y$, with isotropy group $C_3$.\\

III: ${I_4}^3\neq 6{I_6}^2$, $3I_2{I_4}^2 - 6I_4I_6 + 4{I_2}^2 I_6) = 18I_{10}\neq 0$, $4{I_4}^3 +5I_2I_4I_6 +6{I_6}^2 = 3 I_2 I_{10}$.\\ 
Cuspidal curve $(y - 2 + 3x)^2 = 4(1-x)^3$, 
$\Big( z + (1-x)(3x-4) \Big)^2 =  (4 - x)^2 (1 - x)^3$, $y^3\neq x^2$, isotropy $C_2$.\\

II: Requires $I_2 = 0$. This does not occur in our orbit space.\\ 

I:  The boundary surface with the exception of the cusps and cuspidal curves. Neither Bolza, nor Clebsch, included an explicit expression for this surface. We gave several implicit and parametrized expressions in the previous section.\\

Finally, missing from Bolza's list is the case III $\cap$ IV which corresponds to the cusp where the isotropy group is $C_\infty$.\\

It is of some interest to find the locus where the discriminant of the binary sextic vanishes, implying coinciding roots. This discriminant condition is given in terms of our, i.e. Clebsch's, invariants by \cite{Me}
\begin{equation}
\frac{2^6}{5^5} {I_2}^5 - \frac{2^3}{5^2} {I_2}^3 I_4 + I_2 {I_4}^2 - \frac{2^3}{15} {I_2}^2 I_6 + 2 I_4 I_6 + \frac32 I_{10} \ = \ 0 
\end{equation}
Translated into our initial orbit space coordinates (\ref{original-xyz}), this becomes
\begin{equation}
 \frac{2^{8} 3^3}{5^5} - \frac{2^4 3^2}{5^2} x + 3 x^2 
  - \frac{8}5 y +  xy + z \ = \ 0
\end{equation}
In terms of the tangent developable coordinates $x, y, w$  this reads
\begin{equation}
18750 w = (96 - 125 x)(1125 x + 625 y - 864)
\end{equation}
This is a hyperbolic paraboloid which touches, but does not intersect, the boundary surface of the orbit space for $t\in [0,\frac14]$ along the curve
\begin{equation}
x = \frac{16}{25} (1 + 5t^2) \ , \ 
{\bf y} = \frac{64}{125} (1 - 30t^2 - 25t^3) \ , \ 
w = \frac{128}{9375} (1 - 25t^2)(11 - 375t^2 - 500t^3) 
\end{equation}
On this curve, which runs from the point where the isotropy group is $C_\infty$ to the point $(\frac{21}{25},-\frac{81}{125},\frac{486}{3125})$ on the $C_2$ cuspidal edge, the pattern of zeroes of the binary sextic is $(2,2,1,1)$, i.e. two pairs of roots coincide. At the cusp  $C_\infty$ we have the root pattern $(3,3)$. Other degenerate root patterns are not possible for $SU(2)$ binary forms.\\

The above shows that we have described, in section 3, also the  geometry of (a real section of) the moduli space of hyperelliptic curves of genus two. We are not aware of such results in the mathematical literature. Furthermore, we have shown that the boundary of this moduli space is a non-orientable tangent developable surface.\\

Recently, mathematicians have started work on the moduli space of hyperelliptic curves of genus three \cite{LeRi}. We expect that, for a suitable choice of coordinates, also this moduli space will have the structure of a (generalized) developable hypersurface.

\section*{Conclusions}
We have studied the geometry of the orbit space for several representations of $SO(3)$ which have in common that there exists a single polynomial relation among the invariants. The strata of the orbit space for an octupole tensor were shown to consist of three cusps, two cuspidal curves, a boundary surface, and its interior. We have found that an essentially unique choice of orbit coordinates turns this boundary surface into a tangent developable of degree six. Being developable, the surface has zero Gaussian curvature. This means in particular that, after dissecting the surface along its cuspidal curves, the three pieces can be isometrically mapped to the plane. We identified this sextic developable surface with one in the classification by Chasles and Cayley. For a set of three vectors, the orbit space was shown to have the Cayley nodal cubic surface as its boundary. The orbit space for a vector plus quadrupole was found to have as its boundary Cremona's first species in his list of quartic ruled surfaces. For these reducible representations then, the boundary of the corresponding orbit spaces is not a developable surface.\\

We have argued that the orbit space of a real traceless symmetric rank-$\ell$ tensor of $SO(3)$ can be identified with the moduli space of an $SU(2)$ binary form of degree $2\ell$. Hence, our description of the orbit space of an octupole also yields the geometry of the moduli space of an $SU(2)$ binary sextic and hence of the moduli space of hyperelliptic curves of genus two. We have shown that the boundary surface of this moduli space is a non-orientable tangent developable. We expect the boundary of the moduli space of higher genus hyperelliptic curves to be generalized developable hypersurfaces.\\

These results have led us to investigate all three-dimensional orbit spaces for irreducible representations of the classical and exceptional Lie groups studied in \cite{K}. In all cases we found that an essentially unique choice of the orbit space coordinates makes the boundary a tangent developable surface\footnote{To be published elsewhere. For the adjoint of $F_4$, only part of the surface can be made developable, the remainder being a non-developable ruled surface but the r\^oles can also be reversed.}. We conclude that the boundary of a three-dimensional orbit space of an otherwise arbitrary irreducible representation of a Lie group is a tangent developable surface. Clearly, a further understanding of the cause of the developable nature of all these spaces is needed.\\

Possible extensions of our work, some of it now in progress, include: i) Study the geometry of the orbit spaces of all non-coregular $SL(2,C)$ representations with a single syzygy; ii) Study the geometry of the complete orbit space of an $\ell = 4$ tensor of $SO(3)$. The structure of the ring of invariants of a binary octavic is known and it has three levels of syzygies. It should be interesting to see how these relations determine the shape of this orbit space. Recent work by \cite{LeRi} looks at the relevance for hyperelliptic curves of genus three ; 
iii) Study the geometry of $n$-dimensional orbit spaces where $n\geq 4$. We expect these to be generalized ruled spaces; iv) Study canonical forms for $SO(3)$ tensors of order higher than three. These tensors typically have several associated quadrupole tensors. The question then arises which of these quadrupoles should dictate the canonical form of its parent tensor. A natural choice is to take the non-degenerate quadrupole tensor of lowest order.\\ 

\section*{Appendix: Cubic Multigraphs and Octupole Invariants}
The a priori number of invariants of order $2n$ of an $\ell = 3$ tensor equals the number of connected cubic multigraphs without loops with $2n$ vertices. The chemist A.T. Balaban studied such graphs already in the nineteensixties; see Fig. 4 of \cite{BB} for a picture of the 71 planar such graphs at order ten and \cite{BVMH} for a computer program study. The number of such graphs with $2,4,6,8,10$ vertices is $1,2,6,20,91$ (see The On-Line Encyclopedia of Integer Sequences at http://oeis.org/A000421). However, Invariant Theory tells us that almost all of these are reducible and that only $1,1,1,0,1$ remain as members of a Hilbert basis, plus an order 15 pseudo-scalar. We have programmed the relations between all invariants through order ten. At order two there is just one graph, see Fig. 4. At order four there are two graphs, the second graph in Fig. 4 and the tetrahedron. As already stated, these are related by $I_4 + J_4 = \frac16 {I_2}^2$. At order six there are six graphs.\\
\def\fa{1}
\def\fb{1.9}
\def\fc{4.7}
\def\fd{7.5}
\def\fe{9.8}
\def\ff{12.9}
\begin{figure}[H]
\begin{center}
\begin{tikzpicture}[thick]
\draw[thick] ({\fa},0) circle (0.8 cm);
\draw (\fa+0.566 ,0.566) -- (\fa-0.566 ,0.566);
\draw (\fa-0.773 ,0.207) -- (\fa-0.207 ,-0.773);
\draw (\fa+0.773 ,0.207) -- (\fa+0.207 ,-0.773);
\node at (\fa,-1.2) {$I_6$};

\draw (\fb,-0.7) -- (\fb+0.8 ,-0.7);
\draw (\fb+1.2 ,-0.7) -- (\fb+2 ,-0.7);
\draw[thick] (\fb+1,-0.7) circle (0.2 cm);
\draw (\fb ,-0.7) -- (\fb+1 ,0.866);
\draw (\fb+2 ,-0.7) -- (\fb+1 ,0.866);
\draw (\fb ,-0.7) -- (\fb+1 ,-0.178);
\draw (\fb+2 ,-0.7) -- (\fb+1 ,-0.178);
\draw (\fb+1 , 0.866) -- (\fb+1,-0.178);
\node at (\fb+1,-1.2) {$J_6$};

\draw[thick] (\fc,0) circle (0.7 cm);
\draw (\fc,0.7) -- (\fc,-0.7);
\draw (\fc+0.7,0) -- (\fc+1.1 ,0);
\draw[thick] (\fc+1.8,0) circle (0.7 cm);
\draw (\fc+1.8,0.7) -- (\fc+1.8,-0.7);
\node at (\fc+0.9,-1.2) {$K_{6}$};

\draw (\fd,0.7) -- (\fd+2,0.7);
\draw (\fd,-0.7) -- (\fd+2,-0.7);
\draw (\fd,0.7) -- (\fd,-0.7);
\draw (\fd+2,0.7) -- (\fd+2,-0.7);
\draw (\fd+0.5,0) -- (\fd+1.5,0);
\draw (\fd+0.5,0) -- (\fd,-0.7);
\draw (\fd+1.5,0) -- (\fd+2,-0.7);
\draw (\fd+0.5,0) -- (\fd,0.7);
\draw (\fd+1.5,0) -- (\fd+2,0.7);
\node at (\fd+1,-1.2) {$P_6$};

\draw (\fe,0.7) -- (\fe+2,0.7);
\draw (\fe,-0.7) -- (\fe+2,-0.7);
\draw (\fe,0.7) -- (\fe,-0.7);
\draw (\fe+2,0.7) -- (\fe+2,-0.7);
\draw (\fe+0.5,0) -- (\fe+1.5,0);
\draw (\fe+0.5,0) -- (\fe,-0.7);
\draw (\fe+1.5,0) -- (\fe+2,-0.7);
\draw (\fe,0.7) -- (\fe+1.5,0);
%\draw (\fe+2,0.7) -- (\fe+0.5,0);
\draw (\fe+2,0.7) -- (\fe+1.07,0.266);
\draw (\fe+0.5,0) -- (\fe+0.92,0.196);
\node at (\fe+1,-1.2) {$Q_6$};

\draw[thick] (\ff,0) circle (0.8 cm);
\draw (\ff-0.8,0) -- (\ff+0.8,0);
\draw (\ff-0.69,0.4) -- (\ff+0.69,0.4);
\draw (\ff-0.69,-0.4) -- (\ff+0.69,-0.4);
\node at (\ff+0.2,-1.2) {$R_6$};
\end{tikzpicture}
\end{center}
\caption{Graphical representation of all invariants of degree six.}
\end{figure}
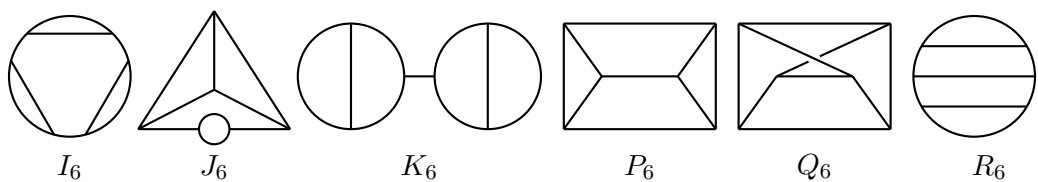

These invariants can all be expressed in terms of lower invariants and $I_6$ as follows
\begin{equation}
J_6 = \frac12 K_6 = \frac16 I_2 I_4 +I_6\ ,\ 
P_6 = \frac1{36} {I_2}^3 + I_6 \ ,\ 
Q_6 = \frac1{18} {I_2}^3 + I_2 I_4 - I_6 \ ,\  
R_6 = \frac16 I_2 I_4
\nonumber
\end{equation}
Note that $R_6$ is reducible, hence it cannot take the place of $I_6$ in the Hilbert basis. At order eight there are twenty multigraphs, hence twenty a priori invariants, all of which are reducible. At order ten there are 91 multigraphs but we find that only 35 of the corresponding invariants are irreducible. Clearly, some care is needed in choosing the single member of order ten of the Hilbert basis.
% N. 9-06-2017

\end{document}